\documentclass[twoside,fleqn]{article}

\usepackage{feynmp}

\usepackage{espcrc2}
\usepackage{epsfig}
\newcommand{\cO}{\mathcal{O}} 
\newcommand{\Dd}[1]{\mbox{
  \parbox[b]{0cm}{$D$}\raisebox{1.7ex}{$\leftrightarrow$}$_{\!#1}$}}

\newcommand{\AmS}{{\protect\the\textfont2
  A\kern-.1667em\lower.5ex\hbox{M}\kern-.125emS}}

\begin{document}
 
%declarations for front matter

\title{Four--quark operators in hadrons\thanks{Talk presented by 
       M. G\"ockeler.}}

\author{
S. Capitani\address{MIT, Center for Theoretical Physics, LNS, 
   77 Massachusetts Avenue, Cambridge, MA 02139, USA},
M. G\"ockeler\address{Institut f\"ur Theoretische Physik,
     Universit\"at Regensburg, D-93040 Regensburg, Germany},
R. Horsley\address{John von Neumann--Institut f\"ur Computing NIC,
        D-15735 Zeuthen, Germany}%
$^,$\address{Institut f\"ur Physik, Humboldt-Universit\"at zu Berlin,
       D-10115 Berlin, Germany},
B. Klaus\address{Institut f\"ur Theoretische Physik,
    Freie Universit\"at Berlin, D-14195 Berlin, Germany}$^{\rm ,c}$,
W. K\"urzinger$^{\rm e,c}$,
D. Petters$^{\rm e,c}$,
D. Pleiter$^{\rm e,c}$,
P.E.L. Rakow$^{\rm b}$,
S. Schaefer$^{\rm b}$,
A. Sch\"afer$^{\rm b}$,
and G.~Schierholz$^{\rm c,}$\address{Deutsches Elektronen-Synchrotron DESY,
      D-22603 Hamburg, Germany}}
 
\begin{abstract}
We present calculations of matrix elements of 4--quark operators
in the pion and in the nucleon extracted from quenched Monte Carlo 
simulations at $\beta = 6.0$ using Wilson fermions. These operators are
relevant for higher--twist effects. We are particularly careful to
avoid mixing with lower--dimensional operators by choosing appropriate
flavour structures.
\end{abstract}
 
% typeset front matter (including abstract)
\maketitle
 
\section{INTRODUCTION}

The operator product expansion (OPE) expresses (Nachtmann) moments of hadronic 
structure functions in terms of Wilson coefficients (usually calculated 
in perturbation theory) and nonperturbative hadronic matrix elements. 
The leading contribution in the deep--inelastic limit  $Q^2 \to \infty$
is provided by operators of twist 2 with 
corrections suppressed by powers of $1/Q^2$ coming from operators of 
twist 4 and higher.
Schematically one finds for $n=2,4,6,\ldots$
\begin{equation}  \begin{array}{l} \displaystyle
 \int_0^1 \mathrm dx \, x^{n-2} F_2(x,Q^2) \big|_{\mathrm {Nachtmann}}
\\ \displaystyle
{} =  c^{(2)}_n(Q^2/\mu^2,g(\mu)) A^{(2)}_n (\mu)
\\ \displaystyle
{} + \frac{c^{(4)}_n(Q^2/\mu^2,g(\mu))}{Q^2} A^{(4)}_n (\mu)
 + O \left(\frac{1}{Q^4}\right) \,.
\end{array}
\end{equation}
The reduced matrix elements $A^{(t)}_n$ of twist $t$ and spin $n$ depend
on the renormalisation scale $\mu$. Whereas the corresponding Wilson 
coefficients $c^{(t)}_n$ are dimensionless, the mass dimension of 
$A^{(t)}_n$ is $t-2$. In the flavour--nonsinglet channel, the twist--2
operators are 2--quark operators,
\begin{equation}
 \bar{\psi}\gamma_{\mu_1} \Dd{\mu_2} \cdots \Dd{\mu_n} \psi \,,
\end{equation}
symmetrised in all indices and with trace terms subtracted.

An important class of twist--4 operators are 4--quark operators.
In particular, the twist--4, spin--2 matrix element $A^{(4)}_2$ is given
by 
\begin{equation} 
  \langle p | A^c_{\{\mu \nu \}} 
     - \mbox{traces} | p \rangle 
 = 2 A^{(4)}_2 (p_\mu p_\nu -  \mbox{traces} ) 
\end{equation}
in terms of the 4--quark operator
\begin{equation} \label{opt4}
 A^c_{\mu \nu } = \bar{\psi} G \gamma_\mu \gamma_5 t^a \psi 
     \bar{\psi} G \gamma_\nu \gamma_5 t^a \psi  \,.
\end{equation}
The quark field $\psi$ carries flavour, colour, and Dirac indices, the
matrices $t^a$ are the usual generators of colour SU(3), and the flavour
matrix $G$ contains the quark charges:
\begin{equation} 
  G = \mbox{diag} ( e_u , e_d ) = \mbox{diag} ( 2/3 , -1/3 ) 
\end{equation}
for two flavours.
The Wilson coefficient reads~\cite{wilco} 
$c_2^{(4)} = g^2 \left( 1+O(g^2) \right)$.

These expressions are to be compared with their twist--2 counterparts:
\begin{equation} 
  \langle p | \cO _{\{\mu \nu \}} 
     - \mbox{traces} | p \rangle 
 = 2 A^{(2)}_2 (p_\mu p_\nu -  \mbox{traces} ) 
\end{equation}
with the operator
\begin{equation} \label{opt2}
  \cO _{\mu \nu} =   
       \frac{\mathrm i}{2} \bar{\psi} G^2 \gamma_\mu \Dd{\nu} \psi 
\end{equation}
and the Wilson coefficient $c_2^{(2)} = 1+O(g^2)$.

The operators (\ref{opt4}) and (\ref{opt2}) transform identically under
Lorentz transformations, but (\ref{opt4}) has dimension 6, wheras 
(\ref{opt2}) has only dimension 4: 4--quark operators will in general
mix with 2--quark operators of lower dimension. This fact complicates
the investigation of 4--quark operators, because the mixing with 
lower--dimensional operators cannot be calculated reliably within
perturbation theory. For the time being, we do not attempt
a nonperturbative calculation of the renormalisation and mixing
coefficients of 4--quark operators. Instead we restrict ourselves
to cases where mixing with lower--dimensional operators is prohibited
by flavour symmetry.

In the following we present Monte Carlo data from quenched simulations
at $\beta = 6.0$ with Wilson fermions on a $16^3 \times 32$ lattice.
From now on all operators are written down in Euclidean space. 

\section{PION}

In the case of the pion, we consider the symmetry SU(2)$_{\mathrm F}$, 
i.e.\ isospin symmetry. While 2--quark operators can have at most
isospin $I=1$, there are 4--quark operators with $I=2$. Such operators
are protected from mixing with lower--dimensional operators and it
makes sense to renormalise them perturbatively. Therefore we study
the $I=2$ pion structure function
\begin{equation}
 F_2^{I=2} = F_2^{\pi^+} + F_2^{\pi^-} - 2 F_2^{\pi^0} \,,
\end{equation}
which is purely higher twist and receives no contributions from
2--quark operators.

Omitting Dirac and colour matrices, the flavour structure of the 
relevant operator in the OPE is given by
\begin{displaymath} \begin{array}{l} \displaystyle
 (e_u \bar{u} u + e_d \bar{d} d )(e_u \bar{u} u + e_d \bar{d} d )
\\ \displaystyle {}
  = \frac{1}{6}  [   
 (\bar{u} u)(\bar{u} u) + (\bar{d} d)(\bar{d} d) 
\\ \displaystyle {}
  - (\bar{u} u)(\bar{d} d) - (\bar{d} d)(\bar{u} u)
 - (\bar{u} d)(\bar{d} u) - (\bar{d} u)(\bar{u} d) ]
\end{array}
\end{displaymath}
up to contributions with $I=0,1$.
Although this flavour structure excludes mixing with 2--quark operators,
mixing with other 4--quark operators is still possible. 
For spin 2, a basis of 
4--quark operators whose $I=2$ components are closed under 
renormalisation is given by the spin--2 projections of 
\setlength{\arraycolsep}{0.5ex}
\begin{equation} \begin{array}{rcl} \displaystyle
 V^c_{\mu \nu } & = & \bar{\psi} G \gamma_\mu t^a \psi 
     \bar{\psi} G \gamma_\nu  t^a \psi   \\
  A^c_{\mu \nu } &  = & 
     \bar{\psi} G \gamma_\mu \gamma_5 t^a \psi 
        \bar{\psi} G \gamma_\nu \gamma_5 t^a \psi  \\
 T^c_{\mu \nu } & = & \bar{\psi} G \sigma_{\mu \rho} t^a \psi 
     \bar{\psi} G \sigma_{\nu \rho} t^a \psi 
\end{array}
\end{equation} 
together with the operators $V_{\mu \nu }$, $A_{\mu \nu }$, and
$T_{\mu \nu }$ which differ from the above by the omission of the colour
matrices $t^a$. 
One--loop results for the renormalisation coefficients are 
given in \cite{pion}. These are to be combined with the bare matrix
elements extrapolated to the chiral limit. Expressing the matrix elements
in terms of the pion decay constant $f_\pi$ we get~\cite{pion}
\begin{equation}
A_2^{(4)\,I=2} = 0.133(51)\, f^{\,2}_{\pi} \,,
\end{equation}
where $f_{\pi} = 131$MeV in the real world. The order of magnitude of
this result is reproduced by the vacuum insertion approximation, which
yields an answer proportional to $f^{\,2}_{\pi}$.
For the first moment of $F_2$ we find
\begin{equation} \label{pionres}
\begin{array}{l} \displaystyle
 \int_0^1 \mathrm dx \,  F_2(x,Q^2) \big|_{\mathrm {Nachtmann}}^{I=2}
\\ \displaystyle {}
 =  1.67(64)\, \frac{f_{\pi}^{2}\,\alpha_s(Q^2)}{Q^2} + O(\alpha_s^2) \,.
\end{array}
\end{equation}
Except for very small values of $Q^2$ this is considerably
smaller than the leading twist--2 contribution 
to the $\pi^+$ structure function computed in a quenched Monte Carlo
simulation~\cite{pionrho}: 
\begin{equation}
 \int_0^1 \mathrm dx \, F_2(x,Q^2) \big|_{\mathrm {Nachtmann}}^{\pi^+}
  = 0.152(7) \,.
\end{equation}
Of course, the fact that the particular twist--4 contribution 
(\ref{pionres}) is rather small does not tell us that all other 
higher--twist contributions will be small as well.

\section{PROTON}

In the case of the proton it is more difficult to find 4--quark operators
that are safe from mixing with lower--dimensional operators: the 
expectation value of any $I=2$ operator in the proton vanishes. So we
have to enlarge the flavour symmetry which we consider from 
SU(2)$_{\mathrm F}$ to SU(3)$_{\mathrm F}$, i.e.\ we assume 
three quarks of the same mass. Correspondingly,
the flavour structure of the operator in the OPE is now
\begin{displaymath}
 \cO = (e_u \bar{u} u + e_d \bar{d} d + e_s \bar{s} s)
          (e_u \bar{u} u + e_d \bar{d} d + e_s \bar{s} s) \,.
\end{displaymath}
Whereas 2--quark operators transform under SU(3)$_{\mathrm F}$ according
to $\overline{\mathbf{3}} \otimes \mathbf{3} = \mathbf{1} \oplus \mathbf{8}$,
we have for 4--quark operators: $(\overline{\mathbf{3}} \otimes \mathbf{3})
\otimes (\overline{\mathbf{3}} \otimes \mathbf{3}) 
= 2 \cdot \mathbf{1} \oplus 4 \cdot \mathbf{8} \oplus
    \mathbf{10} \oplus \overline{\mathbf{10}} \oplus \mathbf{27}$ .
4--quark operators with $I=0,1$, $I_3 = 0$, and hypercharge $Y=0$ from the 
$\mathbf{10}$, $\overline{\mathbf{10}}$, $\mathbf{27}$
multiplets do not mix with 2--quark operators and can be used in a 
proton expectation value, e.g.\ the $I=1$ operator
\begin{displaymath} \begin{array}{l} \displaystyle 
\cO^{I=1}_{27} =  \frac{1}{10} [
 (\bar{u} u) (\bar{u} u) - (\bar{d} d) (\bar{d} d) 
\\ \displaystyle  {}
 - (\bar{u} s) (\bar{s} u) - (\bar{s} u) (\bar{u} s) 
 + (\bar{d} s) (\bar{s} d) + (\bar{s} d) (\bar{d} s) 
\\ \displaystyle  {}
 - (\bar{s} s) (\bar{u} u) - (\bar{u} u) (\bar{s} s) 
 + (\bar{s} s) (\bar{d} d) + (\bar{d} d) (\bar{s} s) ]
\end{array}
\end{displaymath}
belongs to the $\mathbf{27}$ multiplet. Analogously to the case of the pion,
one can find linear combinations of the structure functions of
the octet baryons ($p$, $n$, $\Lambda$, $\Sigma$, $\Xi$) which
project out the desired flavour component, e.g.
\begin{displaymath} \begin{array}{l} \displaystyle
 \langle \Sigma^+ | \cO | \Sigma^+ \rangle 
 -2 \langle \Sigma^0 | \cO | \Sigma^0 \rangle 
  + \langle \Sigma^- | \cO | \Sigma^- \rangle 
\\  \displaystyle {}
 = \langle p | \cO^{I=1}_{27} | p \rangle \,, 
\end{array}
\end{displaymath}
\begin{displaymath} \begin{array}{l} \displaystyle
    \langle \Sigma^+ | \cO | \Sigma^+ \rangle 
  + \langle \Sigma^- | \cO | \Sigma^- \rangle 
\\  \displaystyle {}
 +6 \langle \Lambda  | \cO | \Lambda \rangle 
  -2 \langle \Xi^0 | \cO | \Xi^0 \rangle 
  -2 \langle \Xi^- | \cO | \Xi^- \rangle 
\\  \displaystyle {}
 -2 \langle p | \cO | p \rangle 
 -2 \langle n | \cO | n \rangle 
 = - \langle p | \cO^{I=1}_{27} | p \rangle \,.
\end{array}
\end{displaymath}

The proton matrix elements are computed in the standard fashion from 
ratios of 3--point functions $\langle B(t) \cO (\tau) \bar{B}(0) \rangle$
over 2--point functions $\langle B(t) \bar{B}(0) \rangle$ 
($ 0 \ll \tau \ll t $) with suitable interpolating fields $B$ and $\bar{B}$.
For a general 4--quark operator the 3--point function 
$\langle B(t) \cO (\tau) \bar{B}(0) \rangle$ consists of three types
of contributions, which can be represented pictorially by the following
diagrams 
\begin{center}
\begin{fmffile}{fd1}
\begin{fmfgraph}(60,40) %\fmfpen{thick}
\fmftop{i1} \fmfbottom{o1}
\fmf{phantom}{i1,v} \fmf{plain}{v,v}  \fmf{plain,left=90}{v,v} 
\fmf{phantom}{v,o1}
\fmfdot{v} 
\fmfforce{(0.5w,0.8h)}{v}
\fmfforce{(0.1w,0.3h)}{v1}
\fmfforce{(0.9w,0.3h)}{v2}
\fmfv{decoration.shape=circle,decoration.filled=shaded,
      decoration.size=0.1h}{v1}
\fmfv{decoration.shape=circle,decoration.filled=shaded,
      decoration.size=0.1h}{v2}
\fmf{plain,right=.4}{v2,v1} \fmf{plain,left=.4}{v2,v1}
\fmf{plain}{v2,v1} 
\end{fmfgraph} \end{fmffile} 
\begin{fmffile}{fd2}
\begin{fmfgraph}(60,40) %\fmfpen{thick}
\fmfleft{i1}  \fmfright{o1} \fmf{phantom}{i1,v} \fmf{phantom}{v,o1}
\fmfforce{(0.5w,0.54h)}{v}
\fmfforce{(0.1w,0.3h)}{v1}
\fmfforce{(0.9w,0.3h)}{v2}
\fmfv{decoration.shape=circle,decoration.filled=shaded,
      decoration.size=0.1h}{v1}
\fmfv{decoration.shape=circle,decoration.filled=shaded,
      decoration.size=0.1h}{v2}
\fmf{plain,left=.4}{v2,v1} \fmf{plain}{v2,v1} 
\fmf{plain,right=.2}{v2,v,v1}
\fmfdot{v} 
\fmf{plain,tension=1.4}{v,v}
\end{fmfgraph} \end{fmffile} 
\\
\begin{fmffile}{fd3}
\begin{fmfgraph}(60,40) %\fmfpen{thick}
\fmfforce{(0.5w,0.54h)}{v}
\fmfforce{(0.1w,0.3h)}{v1}
\fmfforce{(0.9w,0.3h)}{v2}
\fmfv{decoration.shape=circle,decoration.filled=shaded,
      decoration.size=0.1h}{v1}
\fmfv{decoration.shape=circle,decoration.filled=shaded,
      decoration.size=0.1h}{v2}
\fmf{plain,left=.4}{v2,v1} 
\fmf{plain,left=.4}{v2,v}  \fmf{plain,right=.4}{v2,v} 
\fmf{plain,left=.4}{v,v1}  \fmf{plain,right=.4}{v,v1} 
\fmfdot{v} 
\end{fmfgraph} \end{fmffile} 
\end{center}
It is precisely through contributions of the form of the first two diagrams
that the mixing with lower--dimensional operators occurs. Therefore these
contributions cancel in the operators which we consider, and we are left
with the contributions of the last type only.

\begin{figure}
\epsfig{file=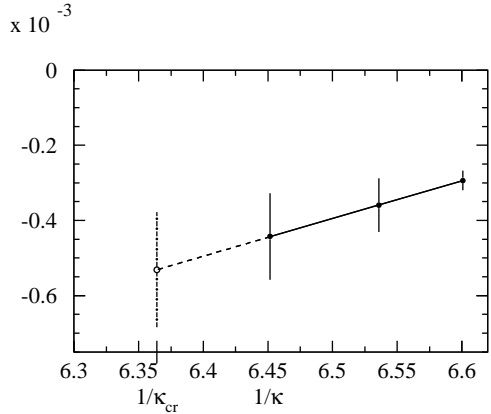,width=7.5cm} 
\vspace*{-1.5cm}
\caption{Chiral extrapolation of the bare proton matrix element of 
$\cO^{I=1}_{27}$ for the operator $V^c_{44} - \mbox{traces}$,
divided by $m_p^4$.}
\label{fig.chiex}
\end{figure}

As an example of our results we show in fig.\ref{fig.chiex} the 
chiral extrapolation of the 
bare proton matrix element of $V^c_{44} - \mbox{traces}$ 
($I=1$ component in the $\mathbf{27}$ representation of 
SU(3)$_{\mathrm F}$) divided by
the fourth power of the proton mass $m_p$. It indicates the order of
magnitude of all 4--quark matrix elements that we have studied.
After renormalisation this 
expression contributes to the reduced matrix element $A_2^{(4)}$ 
according to
\begin{equation} 
 \frac{A^{(4)}_2}{m_p^2} = \frac{2}{3} 
  \frac{\langle N | \cdots | N \rangle}{m_p^4} \,,
\end{equation}
and we obtain for the lowest moment of $F_2$ in our special flavour channel
\begin{equation} \begin{array}{l} \displaystyle
 \int_0^1 \mathrm dx \, F_2(x,Q^2) \big|_{\mathrm {Nachtmann}}
 ^{\mathbf{27}, I=1}
\\ \displaystyle {}
 = - 0.0006(5) \frac{m_p^2 \alpha_s (Q^2)}{Q^2} + O(\alpha_s^2) \,.
\end{array}
\label{nuclres}
\end{equation}
In the proton the corresponding twist--2 contribution is $\approx 0.14$. 
Once again, the twist--4 correction is tiny.
This may be compared with bag model estimates, which give numbers 
$\propto B/m_p^4 \approx 0.0006$ for the prefactor in (\ref{nuclres}),
where $B \approx (145 \mbox{MeV})^4$ is the bag constant~\cite{bag}.

\section{COMPARISON OF RESULTS}

\begin{figure}
\vspace*{-1.2cm}
\epsfig{file=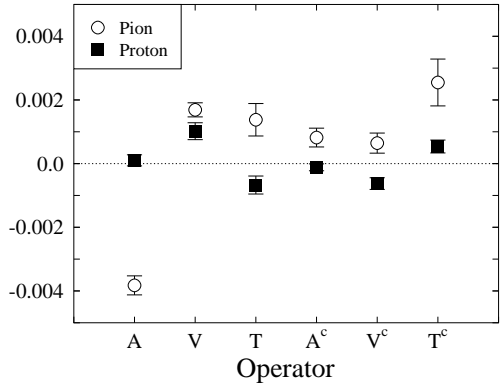,width=8.5cm} 
\vspace*{-1.5cm}
\caption{Renormalised 4--quark matrix elements in the pion and in the proton.}
\label{fig.comp}
\end{figure}

Let us finally compare in fig.\ref{fig.comp} the renormalised 
pion matrix elements
$ \langle \pi^+ | 6 \cdot ( \cdots )^{I=2} | \pi^+ \rangle / m_\pi^2 $
with the corresponding renormalised proton matrix elements 
$ \langle p | 10 \cdot ( \cdots )^{I=1}_{27} | p \rangle / m_p^2 $
(in lattice units).
In the operators we have set $\mu = \nu = 4$ (with traces subtracted).
The normalisation of the pion and proton matrix elements is chosen such that 
the flavour structure $(\bar{u} u) (\bar{u} u)$ appears with the factor 1
in both cases. In view of the fact that pion and proton are very different
particles it is hardly surprising that the numbers do not show many 
similarities.

\section{SUMMARY}

Contributions of higher twist to hadronic structure functions promise
to challenge lattice QCD for a few more years. Still, 
4--quark operators can give reasonable signals in present
quenched Monte Carlo simulations. However, when the flavour structure
is such that it prohibits  mixing 
with lower--dimensional (2--quark) operators, the  matrix elements 
turn out to be rather small. Unfortunately, these flavour structures
are somewhat exotic and not easily accessible to experiment. 
The investigation of physically more 
interesting flavour channels is of course desirable, but impossible
without progress in nonperturbative renormalisation.
It remains to be seen what will happen first: a measurement of structure
functions in the above flavour channels or a reliable lattice calculation
of twist--4 matrix elements relevant for the proton or the pion. 

\section*{ACKNOWLEDGEMENTS}
This work is supported by the DFG (Schwer\-punkt ``Elektromagnetische Sonden'')
and by BMBF. S.C. has been supported in part by the U.S. Department of
Energy (DOE) under cooperative research agreement DE-FC02-94ER40818.
The numerical calculations were performed on the Quadrics 
computers at DESY Zeuthen. We wish to thank the operating staff 
for their support.

%%%%%%%%%%%%%%%%%%%%%%%%%%%%%%%%%%%%%%%%%%%%%%%%%%%%%%%%%%%%%%%%%

\end{document}